\documentclass[aps,prl,twocolumn,superscriptaddress]{revtex4-2}
\usepackage{amsfonts}
\usepackage{amssymb}
\usepackage{amsmath}
\usepackage{epsfig}
\usepackage{color}
\usepackage{graphics, graphicx}
\usepackage{bbold}
\usepackage{psfrag}
\usepackage{mathcomp}
\usepackage{subfigure}
\usepackage{verbatim}
\usepackage{color}
\usepackage[colorlinks,citecolor=blue]{hyperref}

\usepackage{soul}
\usepackage{ulem}
\usepackage{cancel}

\begin{document}

\title{Interaction-induced chiral-transport inversion}

\author{Li Pan}
\thanks{These authors contributed equally to this work}
\affiliation{College of Physics, Sichuan University, Chengdu 610065, China}
\author{Qian Liang}
\thanks{These authors contributed equally to this work}
\affiliation{%
Zhejiang Province Key Laboratory of Quantum Technology and Device, School of Physics, and State Key Laboratory for Extreme Photonics and Instrumentation, Zhejiang University, Hangzhou 310027, China
}%
\affiliation{%
College of Optical Science and Engineering, Zhejiang University, Hangzhou 310027, China
}
\affiliation{Department of Physics, The Pennsylvania State University, University Park, Pennsylvania 16802, USA}
\author{Chang-An Yang}
\affiliation{College of Physics, Sichuan University, Chengdu 610065, China}
\author{Yu Huang}
\affiliation{College of Physics, Sichuan University, Chengdu 610065, China}
\author{Pengjie Liu}
\affiliation{College of Physics, Sichuan University, Chengdu 610065, China}
\author{Fanying Xi}
\affiliation{College of Physics, Sichuan University, Chengdu 610065, China}
\author{Wei Yi}
\email{wyiz@ustc.edu.cn}
\affiliation{Laboratory of Quantum Information, University of Science and Technology of China, Hefei 230026, China}
\affiliation{Anhui Province Key Laboratory of Quantum Network, University of Science and Technology of China, Hefei 230026, China}
\affiliation{CAS Center For Excellence in Quantum Information and Quantum Physics, Hefei 230026, China}
\affiliation{Hefei National Laboratory, University of Science and Technology of China, Hefei 230088, China}
\author{Xiaofan Zhou}
\email{zhouxiaofan@sxu.edu.cn}
\affiliation{State Key Laboratory of Quantum Optics and Quantum Optics Devices, Institute
of Laser spectroscopy, Shanxi University, Taiyuan 030006, China}
\affiliation{Collaborative Innovation Center of Extreme Optics, Shanxi University,
Taiyuan, Shanxi 030006, China}
\author{Jian-Song Pan}
\email{panjsong@scu.edu.cn}
\affiliation{College of Physics, Sichuan University, Chengdu 610065, China}
\affiliation{Key Laboratory of High Energy Density Physics and Technology of Ministry of Education, Sichuan University, Chengdu 610065, China}

\begin{abstract}
{We investigate the chiral dynamics of locally interacting bosons in a two-leg flux ladder, where on-site interactions---despite being fully isotropic---counterintuitively reverse the flux-induced chiral transport of density distribution. For a Bose-Einstein condensate (in the mean-field regime), this reversal arises from an interaction-driven dynamical band-occupation inversion, which selectively populates single-particle states of the opposing chirality.  Strikingly, the chiral-transport inversion has a few-body, hence beyond-mean-field origin,
as the formation of two-body bound states with reversed chirality dominates the few-body dynamics.
This dual pathway---occupation inversion and bound-state formation---underlies the chiral-transport inversion, which challenges the conventional wisdom that isotropic interactions cannot bias density transport.
Our work reveals the interplay between interactions and chirality, and highlights how correlations engineer exotic quantum transport.
}
\end{abstract}

\maketitle

{\it Introduction.---}
{The interplay of interaction and gauge fields plays a key role in a wealth of physical phenomena with distinct contexts and energy scales~\cite{yang1954conservation, jackson2001historical}. For example, the quasiparticles, formed by electrons and fractional magnetic fluxes in two-dimensional electron gases under a magnetic field, lead to the celebrated fractional quantum Hall effect~\cite{tsui1982two, lauphlin1983anomalous}. It is therefore a prominent task to understand the effects of interaction on physical processes induced by gauge fields in both fundamental physics and quantum simulation~\cite{dalibard2011colloquium, cui2013synthetic, galitski2013spin, zhou2013unconventional, celi2014synthetic, goldman2014light, zhai2015degenerate, yi2015pairing, lin2011spin, meng2016experimental, huang2016experimental, wu2016realization, kolkowitz2017spin}.

\begin{figure*}
  \centering
  \includegraphics[width=17cm]{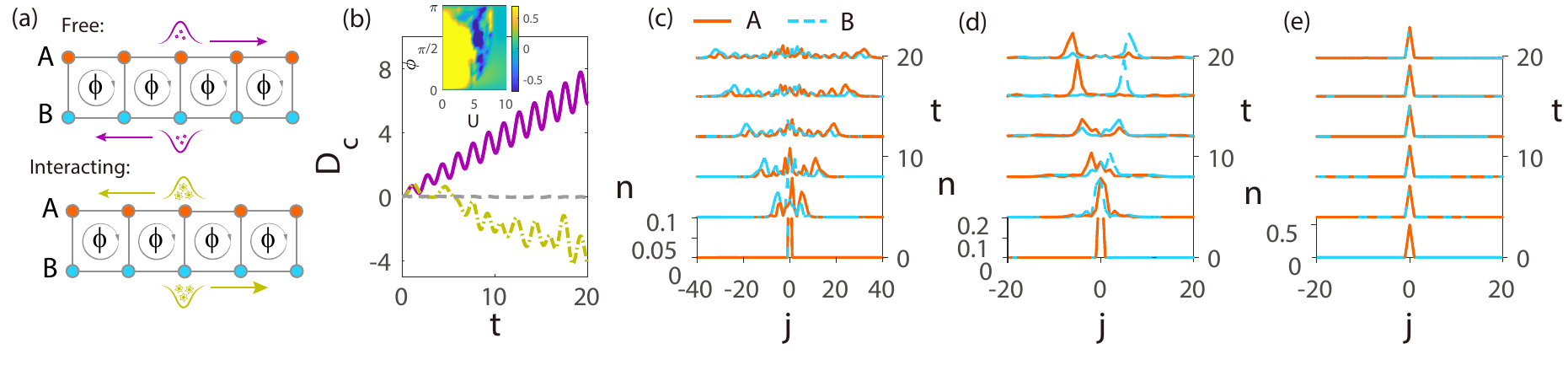}\\
  \caption{Chiral transport of a BEC. (a) Illustration of the IIICC phenomenon in a square flux ladder. As illustrated in the lower panel, interactions induce the inversion of chiral transport. (b) Chiral displacement under different interactions: $U=0$ (purple solid), $U=5$ (laurel green dash-dotted), and $U=10$ (gray dashed). Inset: phase diagram where the color contour represents the averaged chiral displacement over the evolution time $t_{f}=20$.  (c)-(e): the evolution of wave functions for $U=0$ (c), $U=5$ (d) and $U=10$ (e), respectively. The magenta solid and lake blue dashed curves represent the components A and B, respectively. For the calculations, we take $\phi=\pi/2$, and the length of the ladder $L=101$. The BEC is initially prepared in a superposition state $|\psi_{ini}\rangle$ [see Eq.~(\ref{eq:initial_state})]. The units of length, energy and time in this paper are fixed by setting J=1 and the lattice constant as $1$.}\label{fig:PD_MF}
\end{figure*}

Ultracold atoms provide an ideal platform for studying the effects of gauge field, as well as the impact of interactions. For instance, the ground-state quantum phase transition from a Meissner to a vortex phase~\cite{meissner1933ein, hardeen1957theory, khomski1994wohlleben, geim1998paramagnetic, orignac2001meissner, hirsch85the, petrescu2013bosonic, silva2015giant} for bosonic square flux ladders~\cite{hugel2014chiral} was predicted and subsequently experimentally observed in cold atoms~\cite{atala2014observation}. In either phase, atoms flow in opposite directions along different legs of the ladder, with chirality determined by the direction of the flux, akin to the chiral edge currents of a quantum Hall system~\cite{klitzing1980new, thouless1982quantized, aidelsburger2013realization, miyake2013realizing, aidelsburger2015measuring}. {Recent advances in quantum simulation have unveiled rich interplay between interactions and synthetic gauge fields, ranging from {the observation of universal Hall responses in interacting fermions~\cite{zhou2023observation}}  and interaction-driven fractional quantum Hall states in few-body systems~\cite{leonard2023realization}, to scalable realizations of chiral Mott-Meissner phases in bosonic ladders~\cite{impertro2024strongly}. These works highlight how strong correlations can coexist with or even enhance topological phenomena, offering new pathways to engineer exotic quantum matter.}

Chiral behavior also emerges in the dynamics following a localized initial state in the presence of synthetic gauge field, where the chiral transport of time-dependent particle distribution is considered, rather than chiral currents with static particle distributions of the ground state~\cite{tai2017microscopy}.
While chiral transport is generally sensitive to the ladder geometry and distinct forms of the synthetic gauge field, it is also affected by interactions. For instance, the long-range interactions unique to a synthetic momentum lattice suppress the chiral flow, giving rise to self-trapping~\cite{ozwa2015momentum, an2017direct, an2018correlated, li2023observation, liang2023chiral, li2024engineering}.

In this work, we theoretically study the impact of on-site repulsive interactions on the chiral transport of density distribution in a bosonic square flux ladder, inspired by an experimental work in which the chirality of the dynamics of a local two-particle initial state is shown to be switched on by interactions~\cite{tai2017microscopy}. The essential difference is that the many-body initial state is set as a superposition state here, rather than a direct product state studied previously. We find that the isotropic contact interaction counterintuitively reverses the direction of the chiral flow starting from a localized initial state.
This result holds across various correlation regimes.
In the mean-field regime, where the dynamics of the BEC is captured by the Gross-Pitaevskii (GP) equation, such an interaction-induced chiral-transport inversion (IICTI) is attributed to the interaction-induced dynamic population of the single-particle band with opposing chirality.
Beyond the mean-field regime, we show that the chirality inversion
also emerges in the few-body dynamics, using both the time-dependent density-matrix renormalization group (t-DMRG) and exact diagonalization (ED) analyses.
Similar to the mean-field case, the occupation of scattering states with different chirality still contributes to the observed IICTI.
But, intriguingly,  the dynamic formation of two-body bound state with reversed chirality also
has a significant contribution to the IICTI.
This is corroborated by the evolution of the two-body correlation, which exhibits opposite chirality compared to single-particle transport.
Our results reveal novel mechanisms under which chiral dynamics can be tuned by interactions,
as well as the significance of correlation effects in quantum transport,
}

{\it Model.---}
We concentrate on a square flux ladder, wherein a constant flux penetrates each square plaquette [see Fig.~\ref{fig:PD_MF}(a)]. The Hamiltonian is expressed as $H=H_{0}+H_{\text{int}}$, where
\begin{align}
H_{0}&=-J\sum_{j}(e^{-i\frac{\phi}{2}}\hat{a}_{j}^{\dagger}\hat{a}_{j+1}+e^{i\frac{\phi}{2}}\hat{b}_{j}^{\dagger}\hat{b}_{j+1}+\hat{b}_{j}^{\dagger}\hat{a}_{j}+ \text{H.c.})\\
H_{\text{int}}&=U\sum_{j\xi=a,b}\hat{n}_{j,\xi}(\hat{n}_{j\xi}-1),
\label{eq:HH_Ham_free}
\end{align}
with $\hat{a}_{j}$ and $\hat{b}_{j}$ denoting the annihilation operators for bosons on the $j$-th sites of legs A and B, respectively, and $\hat{n}_{j\xi}=\hat{\xi}_{j}^{\dagger}\hat{\xi}_{j}$ for $\xi=a,b$.
{Note that similar models have been experimentally implemented using ultracold atoms in optical lattice potentials subject to Raman-laser-assisted synthetic flux~\cite{atala2014observation,stuhl2015visualizing, mancini2015observation, livi2016synthetic, tai2017microscopy, kennedy2015observation, wang2024realizing, impertro2024strongly}.}
Alternatively, the model can also be engineered in the synthetic dimensions of a momentum lattice~\cite{ozwa2015momentum, an2017direct, an2018correlated, li2023observation}.
{Here we mainly focus on the on-site interaction with tunable strength $U$.} {We also note that on-rung interactions $H_{\text{int}}=U_{ab}\sum_{j}\hat{n}_{j,a}\hat{n}_{j,b}$, which we neglect here, also lead to a reversal of the chiral transport, although the signature is less pronounced (see the Supplemental Material~\cite{SM} for the phase diagram and more discussions).}

{The flux ladder~\cite{hugel2014chiral} considered is an ideal model for studying the interplay between interaction and gauge fields. On one hand, it is experimentally accessible with cold atoms, where the flexible control over synthetic flux, synthetic ladder geometry, and interaction provides rich possibilities for quantum simulation~\cite{atala2014observation, stuhl2015visualizing, mancini2015observation, livi2016synthetic, tai2017microscopy, an2017direct, kennedy2015observation, wang2024realizing,impertro2024strongly}. On the other hand, it is a minimal setup where the dynamic orbital effects of the synthetic magnetic fields can be investigated using a variety of complementary theoretical approaches~\cite{grechner2015spontaneous, kolley2015strongly, natu2015bosons, uchino2016analytical, greschner2016symmetry, zhou2017interaction, citro2018quantum, buser2019finite, buser2020interacting, qiao2021quantum, palm2021bosonic, zheng2023two, huang2024spatial, giri2023flux, giri2024flux}.}

We focus on the chiral transport starting from a local initial state. The chirality is quantified by the chiral displacement~\cite{maffei2018topological}
\begin{equation}\label{eq:chiral_basied_density}
\hat{D}_{\text{c}}=\sum_{j}j(\hat{a}^\dagger_{j}\hat{a}_{j}-\hat{b}^\dagger_{j}\hat{b}_{j}).
\end{equation}
This operator measures the net chiral transfer of probability density along the two-leg ladder. Specifically, transport along the positive (negative) direction of leg A (B) is considered positive, consistent with the definition of chirality. For a given state $|\psi\rangle$, the horizontal chiral displacement of the center of mass is given by $D_{\text{c}}=\langle\psi|\hat{D}_{\text{c}}|\psi\rangle$.

For the initial state, we consider either a condensate (coherent state) or a two-particle Fock state on the local superposed single-particle mode
\begin{equation}\label{eq:initial_state}
|\psi_{\text{ini}}\rangle=\frac{\sqrt{2}}{2}(\hat{a}_{0}^{\dagger}+\hat{b}_{0}^{\dagger})|0\rangle.
\end{equation}
This superposition state has no classical counterpart, which is a crucial factor contributing to the IICTI phenomenon observed in this study~\cite{initialstate}.
The dynamical evolution of the system starting from this type of condensate (two-particle) state is analyzed using the mean-field method (t-DMRG and ED methods) due to the negligible (non-negligible) quantum fluctuations. It is worth emphasizing that a different two-particle initial state is adopted
in the existing literature~\cite{tai2017microscopy, giri2023flux, giri2024flux} with similar setups, with $|\psi_{\text{ini}}^{(2)}\rangle=\hat{a}_{0}^{\dagger}\hat{b}_{0}^{\dagger}|0\rangle$.

{\it Chiral transport of a BEC.---}
The dynamic evolution of the BEC on the ladder can be analyzed with the mean-field theory. The starting point of the mean-field analysis is the GP equation
\begin{equation}\label{eq:GP_equation}
i\partial_{t}\psi_{\lambda}=\sum_{\rho}H_{\lambda\rho}\psi_{\rho}+2U|\psi_{\lambda}|^{2}\psi_{\lambda},
\end{equation}
where $\psi_{\lambda=2(j-1)+1}=\alpha_{j}=\langle\hat{a}_{j}\rangle$, and $\psi_{\lambda=2(j-1)}=\beta_{j}=\langle\hat{b}_{j}\rangle$. Here the odd- and even-indexed elements of the vector $\psi$ represent the wave functions on leg A and B, respectively. And $H_{\lambda\rho}$ are the corresponding matrix elements of the single-particle Hamiltonian. The GP equation is solved iteratively.

Initially, the condensate is prepared in the local superposition mode $|\psi_{\text{ini}}\rangle$ with zero initial chiral displacement. As illustrated in Fig.~\ref{fig:PD_MF}(b), for typical values of flux and interaction strength, while the noninteracting chiral displacement remains positive (the purple solid curve), the chiral displacement under interactions can turn negative (the laurel green dash-dotted curve). Further increase in the interaction strength ultimately inhibits the chiral transport (the gray dashed curve), indicating self-trapping. The inset of Fig.~\ref{fig:PD_MF}(b) indicates different dynamic regimes from the mean-field calculations, where the background color represents the time-averaged chiral displacement. When $U=0$, the averaged chiral displacement is positive, consistent with the direction of flux. However, as $\phi$ exceeds a certain threshold, a blue region emerges, signifying the interaction-induced reversal of chiral transport.

In Figs.~\ref{fig:PD_MF}(c)-(e), we show snapshots of the modulus of the wave function at different times.
In the noninteracting case shown in Fig.~\ref{fig:PD_MF}(c), although the wave function expands in both directions, the component on leg A is dominant in the positive direction, while the component on leg B dominates in the negative direction. This results in positive chiral transport. Conversely, as depicted in Fig.~\ref{fig:PD_MF}(d) with $U=5$, the wave-function evolution clearly exhibits opposite chirality. When the interaction strength is sufficiently strong, the condensate remains localized at all times, as seen in Fig.~\ref{fig:PD_MF}(e). These observations are consistent with the chiral-displacement calculations, and provide clear evidence that the interaction can reverse the direction of the chiral transport.

{\it Stroboscopic projection.---}
The interaction-induced inversion of chiral transport is particularly mysterious since the interaction term commutes with the chiral displacement operator, with $[H_{\text{int}},\hat{D}_{\text{c}}]=0$.
However, since $[H_{0},H_{\text{int}}]\neq 0$, the interaction term can affect the dynamics of $\hat{D}_{\text{c}}$ by mixing the eigenstates of $H_{0}$. It implies that the interaction-induced transport inversion is a typical correlated phenomenon.

Indeed, the impact of interactions is more visible by stroboscopically decomposing the dynamic evolution.
Specifically, for an infinitesimally small time step $\Delta t$,
we have $\exp[-iH\Delta t/\hbar]\approx \exp(-iH_{0}\Delta t/\hbar)\exp(-iH_{\text{int}}\Delta t/\hbar)$. Given that $[H_{\text{int}},\hat{D}_{\text{c}}]=0$, the quantity $D_{\text{c}}$ remains invariant during the evolution time step governed by $H_{\text{int}}$ alone. Consequently, any changes in $D_{\text{c}}$ due to interactions can only be attributed to the modification in the stroboscopic projection of the time-evolved state onto the eigenstates of $H_{0}$.

\begin{figure}
  \centering
  \includegraphics[width=8.7cm]{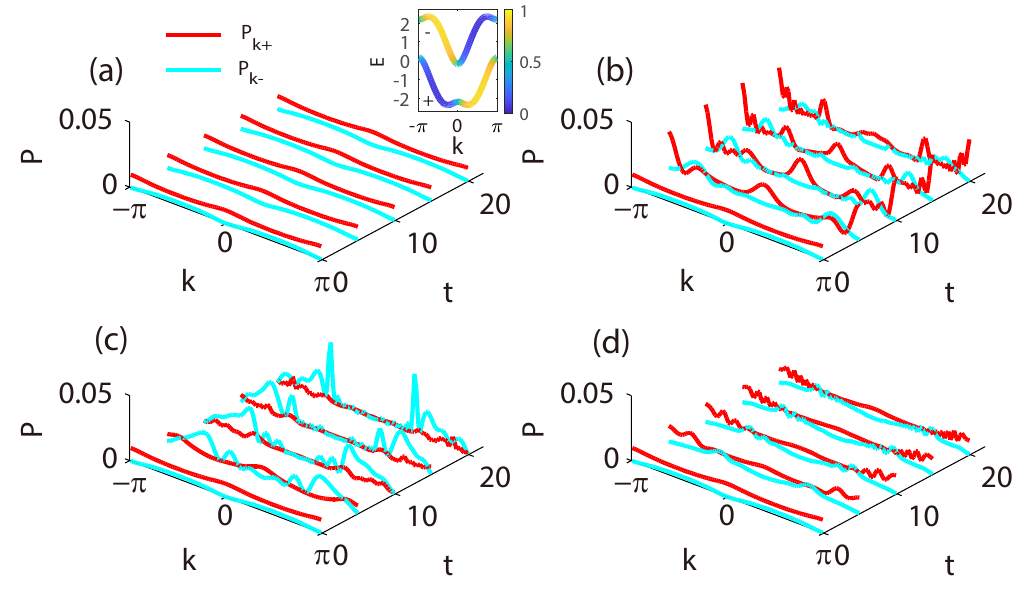}\\
  \caption{Stroboscopic projections $P_{k\pm}$ of the time-evolved wave function onto the Bloch states of the free Hamiltonian. (a)-(d) depict the stroboscopic projection probabilities for cases with $U=0, 2.5, 5$, and $10$, respectively. The inset in (a) displays the energy bands, with the background color representing the magnitudes of component A of the Bloch states. The red and cyan curves represent projections onto the lower ($P_{k+}$) and upper ($P_{k-}$) bands, respectively. In the case with $U=0$, the projection probabilities remain constant over time and are identical to those with finite interaction strengths and at $t=0$. Notably, in (c) where an intermediate interaction is present, a clear inversion of the stroboscopic projection probabilities over time is observed. All other parameters are consistent with those used in Fig.~\ref{fig:PD_MF}.}\label{fig:band_projection}
\end{figure}

In Fig.~\ref{fig:band_projection}, we show the stroboscopic projection of the instantaneous state onto the eigenstates of $H_{0}$, defined as $P_{k\pm}=|\langle\psi(t)|\psi_{k\pm}\rangle|^2$ with the quasimomentum $k\in (-\pi,\pi)$ and the band index $\pm$ labeling the lower ($+$) and upper ($-$)  bands, respectivelyb. The chirality of a noninteracting ladder is primarily governed by the inherent chirality of the energy bands of $H_{0}$~\cite{hugel2014chiral,tai2017microscopy}, where the lower and higher energy bands exhibit quasimomentum-leg locking, as depicted in the inset of Fig.~\ref{fig:band_projection}. Specifically, the eigenstates of the lower (higher) band with positive (negative) quasimomenta have predominant support in leg A [highlighted in yellow in the inset of Fig.~\ref{fig:band_projection}(a)], leading to positive (negative) values of $D_{\text{c}}$ when the occupation of the corresponding energy band is dominant.
From Fig.~\ref{fig:band_projection}, we observe that in scenarios with no interaction (a), with small interaction (b), and with strong interaction (d), the stroboscopic projection $P_{nk}$ onto the lower band (red curves) and higher band (cyan curves) does not exhibit notable inversion. Consistently, $D_{\text{c}}$ does not exhibit inversion in these cases [as indicated by the purple solid and gray dashed curves in Fig.~\ref{fig:PD_MF}(b)].

\begin{figure*}
\centering
\includegraphics[width=16cm]{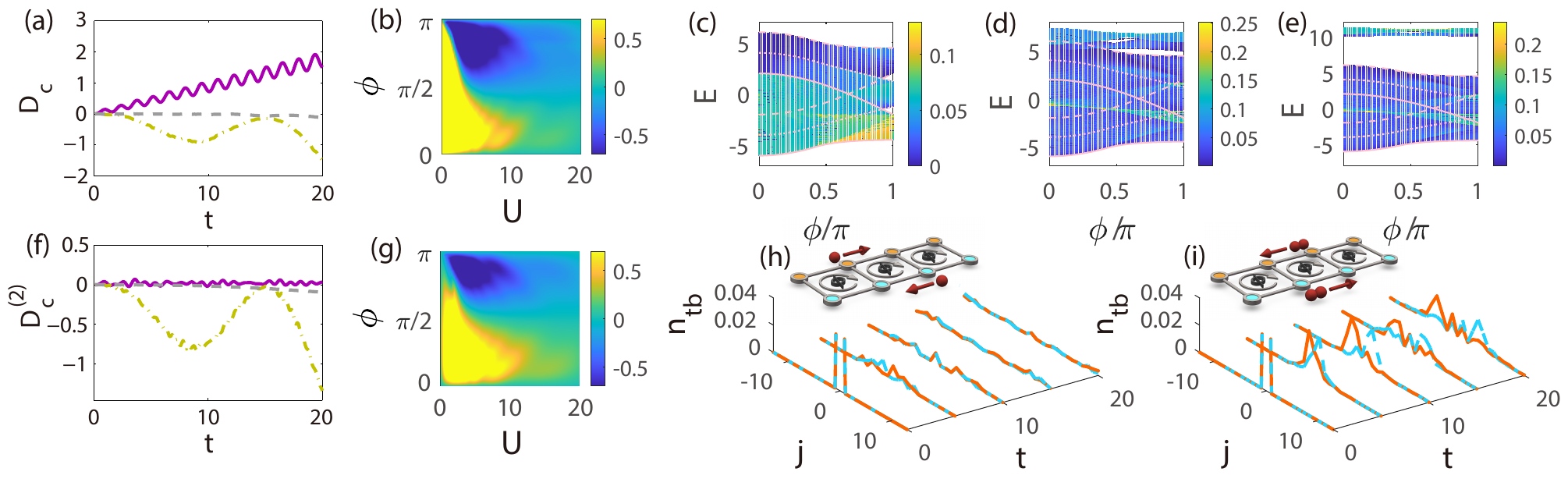}\\
\caption{Chiral transport of a two-body initial state. Chiral displacement $D_{\text{c}}$ (a) and two-particle chiral displacement $D_{\text{c}}^{(2)}$ (f) are plotted for a fixed flux $\phi=0.9\pi$ and different interactions $U=0, 5$ and $10$. The line conventions for different interactions are the same with those in Fig.~\ref{fig:PD_MF}(b).
Subfigures (b) and (g) display the corresponding phase diagrams, where the background colors represent the average values of $D_{\text{c}}$ (b) and $D_{\text{c}}^{(2)}$ (g), respectively. Projection of the two-body initial state onto the eigenstates of the full Hamiltonian given by ED method: (c)-(e) correspond to $U=0,5$ and $10$, respectively. The background color indicates the projection probability. The three scattering bands  defined in the main text, denoted by $|++\rangle$, $|+-\rangle$, and $|--\rangle$, are shown with the pink dash-dotted, dotted and dashed curves, respectively. The evolution of two-particle distribution for $U=0$ (h) $U=5$ (i) are also shown. {The insets illustrate the opposite transports contributed by single-particle and two-particle bound states.} The magenta solid and lake blue dashed curves correspond to the distributions on leg A and B, respectively. The length of the square flux ladder is $41$. Other parameters are the same as those in Fig.~\ref{fig:PD_MF}.}\label{fig:Fig_ED_DMRG}
\end{figure*}

However, for the case with intermediate interaction, as illustrated in Fig.~\ref{fig:band_projection}(c), although initially $P_{nk}$ is dominantly in the lower band (red curves), the projection in the upper band (cyan curve) becomes dominant with increasing evolution time.
This result is consistent with the evolution of $D_{\text{c}}$ [see the laurel green dash-dotted curve in Fig.~\ref{fig:PD_MF}(b)], which initially oscillates near zero but gradually becomes negative with the increase of time. Therefore, the IICTI phenomenon can be attributed to the interaction-induced inversion of stroboscopic projections on the chiral energy bands of the noninteracting Hamiltonian.

{\it Two-particle case.---}
Similar interaction-induced transport inversion is also observed in the few-body case, beyond the mean-field approximation. To see this, we study the chiral transport of two interacting bosons on the ladder.
For this purpose, we employ the t-DMRG method to simulate the dynamic evolution, complemented by studies using the ED.

In Fig.~\ref{fig:Fig_ED_DMRG}(a), we show the results from the t-DMRG calculations. Here $D_{\text{c}}$ exhibits an interaction-induced inversion for intermediate interactions. The corresponding dynamic regimes, represented by the average $D_{\text{c}}$, is shown in Fig.~\ref{fig:Fig_ED_DMRG}(b). Notably, the sign of the chiral displacement is reversed within a similar parameter regime compared to the mean-field case.

We then numerically diagonalize the full Hamiltonian given in Eq.~(\ref{eq:HH_Ham_free}) under the two-particle occupation, and illustrate the projection probabilities of the initial state onto the eigenstates in Figs.~\ref{fig:Fig_ED_DMRG}(c)-(e).
The two-particle scattering eigenstates feature three distinct scenarios, corresponding to three scattering bands: i) both particles occupy the lowest-energy Bloch states on the lower band (denoted by $|++\rangle$); ii) one particle occupies the lower band and the other the upper band ($|+-\rangle$); and iii) both particles occupy the upper band ($|--\rangle$)~\cite{tai2017microscopy}.
The profiles of these three scattering bands are outlined with pink dash-dotted, dotted and dashed curves, respectively. Importantly, the higher occupation of the lowest two-particle band in Fig.~\ref{fig:Fig_ED_DMRG}(c) aligns with the results in Fig.~\ref{fig:band_projection}, leading to a positive chirality $D_{\text{c}}$.

The presence of interaction obscures the boundaries between the scattering bands, as shown in Figs.~\ref{fig:Fig_ED_DMRG}(d) and (e). On one hand, compared to the noninteracting case in Fig.~\ref{fig:Fig_ED_DMRG}(c), the projection probability onto the doubly occupied upper single-particle band $|--\rangle$, which contributes to opposite chiral transport, becomes finite. This observation is consistent with the inversion of stroboscopic projection in the mean-field case. On the other hand, a new band emerges in the presence of interactions [the highest-lying band in Figs.~\ref{fig:Fig_ED_DMRG}(d) and (e)], corresponding to the formation of two-particle bound state~\cite{tai2017microscopy, zheng2023two}.
Notably, the projection probability onto the bound state is even more obvious.

In order to show the chiral displacement induced by the occupation of the two-body bound state, we  compute the evolution of the two-particle chiral displacement, defined as $D_{\text{c}}^{(2)}=\langle\hat{D}^{(2)}_{\text{c}}\rangle$,
where
\begin{equation}\label{eq:tb_chiral_basied_density}
\hat{D}_{\text{c}}^{(2)}=\sum_{j}j(\hat{a}^\dagger_{j}\hat{a}^\dagger_{j}\hat{a}_{j}\hat{a}_{j}-\hat{b}^\dagger_{j}\hat{b}^\dagger_{j}\hat{b}_{j}\hat{b}_{j}).
\end{equation}
The two-particle displacement quantifies the chiral movement of particle pairs.
As illustrated in Figs.~\ref{fig:Fig_ED_DMRG}(f) and (g), the trends observed in $D_{\text{c}}^{(2)}$ mirror those of $D_{\text{c}}$ shown in the same figure.
This similarity suggests that the IICTI phenomenon can be partially attributed to the transport of particle pairs, which can have opposite chirality compared to single particles under the same flux.
A straightforward picture is that particle pairs experience twice the flux and feature opposite chiral transport given the proper range of the flux.
For sufficiently strong interactions, the bound-state bands flatten due to Mott localization,
and no longer contribute to transport. This is confirmed in Figs.~\ref{fig:Fig_ED_DMRG}(f) and (g).

The chiral transport of two-body bound states on a lattice can be analyzed using the degenerate perturbation theory~\cite{takahashi1997half, qin2014statistics, ke2017multiparticle, SM}, where the effective Hamiltonian for the doublon operators
$\hat{A}^{\dagger}_{j}=\hat{a}^{\dagger}_{j} \hat{a}^{\dagger}_{j}$ and $\hat{B}^{\dagger}_{j}=\hat{b}^{\dagger}_{j} \hat{b}^{\dagger}_{j}$ in the strong-interaction limit is
(apart from a uniform on-site energy shift)
\begin{equation}\label{eq:two_body}
H^{(2)}=\frac{2J^2}{U}\sum_{j}(\hat{A}^{\dagger}_{j}\hat{A}_{j+1}e^{-i\phi}+\hat{B}^{\dagger}_{j}\hat{B}_{j+1}e^{i\phi}+\hat{A}^{\dagger}\hat{B}+\text{H.c.}).
\end{equation}
This Hamiltonian represents a flux ladder with a constant flux $\phi^{(2)}=2\phi$. When $\phi^{(2)}>\pi$, the chiral transport of the particle pairs reverses its direction, consistent with the physical picture above.

Finally, we illustrate in Figs.~\ref{fig:Fig_ED_DMRG}(h) and (i) the numerically simulated wave-function evolution of particle pairs.
In the noninteracting case, no discernible chirality is present. Under intermediate interactions, the particle-pair distribution exhibits chirality opposite to that of the single-particle chiral transport in the noninteracting case. This provides further evidence that the IICTI phenomenon in the two-particle case is associated with not only the interaction-induced occupation of scattering bands, but also with the transport of two-particle bound states. It is reminiscent of the transport induced by bound states in a flux lattice exhibiting the Aharonov-Bohm caging effect~\cite{vidal2000interaction, vidal2001disorder}.

{\it Discussions.---}
{We have demonstrated a remarkable reversal of chiral particle transport in a bosonic two-leg flux ladder, driven by isotropic on-site interactions. Our finding defies the conventional wisdom regarding the dynamic impact of local interactions. It is worth noting that the chiral current of ground state may also undergo reversal due to spontaneous symmetry breaking driven by interactions~\cite{grechner2015spontaneous, kolley2015strongly, uchino2016analytical, greschner2016symmetry}. Therein, the underlying mechanism is the interaction-induced spontaneous translational symmetry breaking, which enlarges the unit cell, causing the effective flux to double and match the condition for a current reversal.
However, our work mainly focuses on the inversion of chiral transport of atoms, a dynamic process involving all eigenstates of the Hamiltonian. The underlying mechanism of the phenomenon reported here is not the spontaneous breaking of translation symmetry, and our observation is essentially different from the static chiral current distribution associated with phase gradient of the ground state.
In particular, our results can not be explained by examining the ground state of the system alone, which necessitates the local initial states in our discussion.}


While the setup we consider is readily accessible in cold-atom laboratories, the formation of bound states as a mediator of chiral-transport reversal represents a fundamentally novel mechanism, with potential implications extending beyond the present system. {While we have assumed a zero- temperature condition for the initial condensate and numerical calculations, we expect the presence of thermal atoms to only slightly smear out the signature of chiral transport, given their small fraction and impact under typical experimental conditions~\cite{tai2017microscopy, liang2023chiral}.}
The competition between the single-particle and the two-particle bound states, due to the fact that the two-particle bound states experience a different gauge field, is universal for different types of bound states and gauge fields.
As such, our results not only illuminate a previously overlooked aspect of interaction-driven phenomena, but also pave the way for exploring bound-state-mediated transport dynamics in correlated quantum systems.

{\it Acknowledgments.---}
This work is supported by the National Key R$\&$D Program of China under Grant Nos. 2024YFF0508503 and 2022YFA1404201, the National Natural Science Foundation of China (NSFC) under Grant Nos.~12574297, 12174233, 12004230, 12034012 and 12374479, the Natural Science Foundation of Sichuan Province under Grant No. 2025ZNSFSC0058, the Science Specialty Program of Sichuan University under
Grant No. 2020SCUNL210, the Fundamental Research Funds for the Central Universities under Grant No. YJ202212, and the Fundamental Research Program of Shanxi Province under Grant No. 202403021221024. WY acknowledges support from the Innovation Program for Quantum Science and Technology (Grant No. 2021ZD0301904).


\begin{thebibliography}{99}
\bibitem{yang1954conservation}C. N. Yang, and R. L. Mills,  Conservation of isotopic spin and isotopic gauge invariance, Phys. Rev. \textbf{96}, 191 (1954).

\bibitem{jackson2001historical}J. D. Jackson, and L. B. Okun,  Historical roots of gauge invariance, Rev. Mod. Phys. \textbf{73}, 663 (2001).

\bibitem{tsui1982two}D. C. Tsui, H. L. Stormer, and A. C. Gossard,  Two-Dimensional magnetotransport in the extreme quantum limit, Phys. Rev. Lett. \textbf{48}, 1559 (1982).

\bibitem{lauphlin1983anomalous}R. B. Laughlin, Anomalous quantum Hall effect: an incompressible quantum fluid with fractionally charged excitations, Phys. Rev. Lett. \textbf{50}, 1395 (1983).



\bibitem{dalibard2011colloquium}J. Dalibard, F. Gerbier, G. Juzeli$\ddot{\text{u}}$nas, and P. $\ddot{\text{O}}$hberg, Colloquium: Artificial gauge potentials for neutral atoms, Rev. Mod. Phys. \textbf{83}, 1523 (2011).


\bibitem{cui2013synthetic}X. Cui, B. Lian, T.-L. Ho, B. L. Lev, and H. Zhai, Synthetic gauge field with highly magnetic lanthanide atoms, Phys. Rev. A \textbf{88}, 011601(R) (2013).

\bibitem{galitski2013spin}V. Galitski and I. B. Spielman, Spin-orbit coupling in quantum gases, Nature (London) \textbf{494}, 49 (2013).

\bibitem{zhou2013unconventional}X. Zhou, Y. Li, Z. Cai, and C. Wu, Unconventional states of bosons with the synthetic spin-orbit coupling, J. Phys. B \textbf{46}, 134001 (2013).

\bibitem{celi2014synthetic}A. Celi, P. Massignan, J. Ruseckas, N. Goldman, I. B. Spielman, G. Juzeliunas, and M. Lewenstein, Synthetic Gauge Fields in Synthetic Dimensions, Phys. Rev. Lett. \textbf{112}, 043001 (2014).

\bibitem{goldman2014light}N. Goldman, G. Juzeli$\bar{\text{u}}$nas, P. $\ddot{\text{O}}$hberg, and I. B. Spielman, Light-induced gauge fields for ultracold atoms, Rep. Prog. Phys. \textbf{77}, 126401 (2014).

\bibitem{zhai2015degenerate}H. Zhai, Degenerate quantum gases with spin-orbit coupling: A review, Rep. Prog. Phys. \textbf{78}, 026001 (2015).

\bibitem{yi2015pairing}W. Yi, W. Zhang, and X. Cui, Pairing superfluidity in spin-orbit coupled ultracold Fermi gases, Sci. China: Phys. Mech. Astron. \textbf{58}, 014201 (2015).

\bibitem{lin2011spin} Y.-J. Lin, K. Jim$\acute{\text{e}}$nez-Garcia, and I. B. Spielman, Spin-orbit-coupled Bose-Einstein condensates, Nature (London) \textbf{471}, 83 (2011).

\bibitem{meng2016experimental}Z. Meng, L. Huang, P. Peng, D. Li, L. Chen, Y. Xu, C. Zhang, P. Wang, and J. Zhang, Experimental observation of a topological band gap opening in ultracold Fermi gases with two-dimensional spin-orbit coupling, Phys. Rev. Lett. \textbf{117}, 235304 (2016).

\bibitem{huang2016experimental}L. Huang, Z. Meng, P. Wang, P. Peng, S.-L. Zhang, L. Chen, D. Li, Q. Zhou, and J. Zhang, Experimental realization of two-dimensional synthetic spin-orbit coupling in ultracold Fermi gases, Nat. Phys. \textbf{12}, 540 (2016).

\bibitem{wu2016realization}Z. Wu, L. Zhang, W. Sun, X.-T. Xu, B.-Z. Wang, S.-C. Ji, Y. Deng, S. Chen, X.-J. Liu, and J.-W. Pan, Realization of two-dimensional spin-orbit coupling for Bose-Einstein condensates, Science \textbf{354}, 83 (2016).

\bibitem{kolkowitz2017spin}S. Kolkowitz, S. L. Bromley, T. Bothwell, M. L. Wall, G. E. Marti, A. P. Koller, X. Zhang, A. M. Rey, and J. Ye, Spin-orbit-coupled fermions in an optical lattice clock, Nature (London) \textbf{542}, 66 (2017).



\bibitem{meissner1933ein}W. Meissner, and R. Ochsenfeld,  Ein neuer Effekt bei Eintritt der Supraleitf$\ddot{\text{a}}$higkeit, Naturwissenschaften \textbf{21}, 787-788 (1933).

\bibitem{hardeen1957theory}J. Bardeen, L. N. Cooper, and J. R. Schrieffer, Theory of superconductivity, Phys. Rev. \textbf{108}, 1175-1204 (1957).

\bibitem{khomski1994wohlleben}D. Khomskii,  Wohlleben effect (Paramagnetic Meissner effect) in high-temperature superconductors, Journ. of Low Temp. Phys. \textbf{95} 205-223 (1994).

\bibitem{geim1998paramagnetic}A. K. Geim, S. V. Dubonos, J. G. S. Lok, M. Henini, and J. C. Maan, Paramagnetic Meissner effect in small superconductors, Nature \textbf{396}, 144 (1998).

\bibitem{orignac2001meissner}E. Orignac, and T. Giamarchi, Meissner effect in a bosonic ladder, Phys. Rev. B \textbf{64}, 144515 (2001)

\bibitem{hirsch85the}J. E. Hirsch, The origin of the Meissner effect in new and old superconductors, Phys. Scr. \textbf{85} 035704 (2012).

\bibitem{petrescu2013bosonic}A. Petrescu and K. L. Hur, Bosonic Mott Insulator with Meissner Currents, Phys. Rev. Lett. \textbf{111}, 150601 (2013).

\bibitem{silva2015giant}R. M. da Silva, M. V. Milo$\breve{\text{s}}$evi$\acute{\text{c}}$, A. A. Shanenko, F. M. Peeters, and J. Albino Aguiar, Giant paramagnetic Meissner effect in multiband superconductors, Scientific Reports \textbf{5}, 12695 (2015).


\bibitem{hugel2014chiral}D. H$\ddot{\text{u}}$gel, and B. Paredes, Chiral ladders and the edges of quantum Hall insulators, Physical Review A \textbf{89}, 023619 (2014).




\bibitem{atala2014observation}M. Atala,  M. Aidelsburger,  M. Lohse,  J. T. Barreiro,  B. Paredes, and I. Bloch, Observation of chiral currents with ultracold atoms in bosonic ladders, Nature Physics \textbf{10}, 588--593 (2014).




\bibitem{klitzing1980new}K. v. Klitzing, G. Dorda, and M. Pepper, New Method for High-accuracy determination of the fine-structure constant based on quantized Hall resistance, Phys. Rev. Lett. \textbf{45}, 494 (1980).

\bibitem{thouless1982quantized} D. J. Thouless, M. Kohmoto, M. P. Nightingale, and M. den Nijs, Quantized Hall conductance in a two-dimensional periodic potential, Phys. Rev. Lett. \textbf{49}, 405 (1982).

\bibitem{aidelsburger2013realization}M. Aidelsburger, M. Atala, M. Lohse, J. T. Barreiro, B. Paredes, and I. Bloch, Realization of the Hofstadter Hamiltonian with ultracold atoms in optical lattices, Phys. Rev. Lett. \textbf{111}, 185301 (2013).

\bibitem{miyake2013realizing}H. Miyake, G. A. Siviloglou, C. J. Kennedy, W. C. Burton, and W. Ketterle, Realizing the Harper Hamiltonian with laser-assisted tunneling in optical lattices, Phys. Rev. Lett. \textbf{111}, 199903 (2013).

\bibitem{aidelsburger2015measuring} M. Aidelsburger, M. Lohse, C. Schweizer, M. Atala, J. T. Barreiro, S. Nascimb\`{e}ne, N. R. Cooper, I. Bloch, and N. Goldman, Measuring the Chern number of Hofstadter bands with ultracold bosonic atoms, Nat. Phys. \textbf{11}, 162 (2015).


\bibitem{zhou2023observation}
T.-W. Zhou, G. Cappellini, D. Tusi, L. Franchi, J. Parravicini, C. Repellin, S. Greschner, M. Inguscio, T. Giamarchi, M. Filippone, J. Catani, and L. Fallani,
Observation of universal Hall response in strongly interacting Fermions,
Science \textbf{381}, 427 (2023).

\bibitem{leonard2023realization}
J. L\'{e}onard, S. Kim, J. Kwan, P. Segura, F. Grusdt, C. Repellin, N. Goldman, and M. Greiner,
Realization of a fractional quantum Hall state with ultracold atoms,
Nature \textbf{619}, 495 (2023).






\bibitem{ozwa2015momentum}T. Ozawa, H. M. Price, and I. Carusotto, Momentum-space Harper-Hofstadter model,
Phys. Rev. A \textbf{92}, 023609 (2015).

\bibitem{an2018correlated}F. A. An, E. J. Meier, J. An$\acute{\text{g}}$on$\acute{\text{g}}$a, and B. Gadway, Correlated dynamics in a synthetic lattice of momentum states, Phys. Rev. Lett. \textbf{120}, 040407 (2018).

\bibitem{li2023observation}Y. Li, H. Du, Y. Wang, J. Liang, L. Xiao, W. Yi, J. Ma, and S. Jia, Observation of frustrated chiral dynamics in an interacting triangular flux ladder, Nature Commun. \textbf{14}, 7560 (2023).

\bibitem{liang2023chiral}Q. Liang, Z. Dong, J.-S. Pan, H. Wang, H. Li, Z. Yang, W. Yi, B. Yan, Chiral dynamics of ultracold atoms under a tunable SU (2) synthetic gauge field, Nat. Phys. \textbf{20}, 1738 (2024).

\bibitem{li2024engineering}H. Li, Q. Liang, Z. Dong, H. Wang, W. Yi, J.-S. Pan, and B. Yan, Engineering topological chiral transport in a flat-band lattice of ultracold atoms, arXiv:2401.03611 (2024).

\bibitem{kennedy2015observation} C. J. Kennedy, W. C. Burton, W. C. Chung, and W. Ketterle, Observation of Bose-Einstein Condensation in a Strong Synthetic Magnetic Field, Nat. Phys. \textbf{11}, 859 (2015).

\bibitem{mancini2015observation}M. Mancini, G. Pagano, G. Cappellini, L. Livi, M. Rider, J. Catani, C. Sias,
P. Zoller, M. Inguscio, M. Dalmonte, and L. Fallani, Observation of chiral edge states with neutral fermions in synthetic Hall ribbons, Science \textbf{349}, 1510(2015).

\bibitem{stuhl2015visualizing}B. K. Stuhl, H.-I. Lu, L. M. Aycock, D. Genkina, and I. B. Spielman, Visualizing edge states with an atomic Bose gas in the quantum Hall regime, Science \textbf{349}, 6255 (2015).


\bibitem{livi2016synthetic}L. F. Livi, G. Cappellini, M. Diem, L. Franchi, C. Clivati, M. Frittelli, F. Levi, D. Calonico, J. Catani, M. Inguscio, and L. Fallan, Synthetic dimensions and spin-orbit coupling with an optical clock transition, Phys. Rev. Lett. \textbf{117}, 220401 (2016).


\bibitem{tai2017microscopy}M. E. Tai, A. Lukin, M. Rispoli, R. Schittko, T. Menke, D. Borgnia, P. M. Preiss, F. Grusdt, A. M. Kaufman,  and M. Greiner, Microscopy of the interacting Harper--Hofstadter model in the two-body limit, Nature \textbf{546}, 519--523 (2017).


\bibitem{wang2024realizing} Y. Wang, Y.-K. Wu, Y. Jiang, M.-L. Cai, B.-W. Li, Q.-X. Mei, B.-X. Qi, Z.-C. Zhou, and L.-M. Duan, Realizing Synthetic Dimensions and Artificial Magnetic Flux in a Trapped-Ion Quantum Simulator, Phys. Rev. Lett. \textbf{132}, 130601 (2024).


\bibitem{impertro2024strongly} A. Impertro, S. J. Huh, S. Karch, J. F. Wienand, I. Bloch, and M. Aidelsburger, Strongly interacting Meissner phases in large bosonic flux ladders, Nat. Phys. (2025). https://doi.org/10.1038/s41567-025-02890-0.


\bibitem{an2017direct} F. A.  An, E. J. Meier,  and B. Gadway, Direct observation of chiral currents and magnetic reflection in atomic flux lattices, Sci. Adv. \textbf{3}, e1602685 (2017).

\bibitem{grechner2015spontaneous}S. Greschner, M. Piraud, F. Heidrich-Meisner, I. P. McCulloch, U. Schollw$\ddot{\text{o}}$ck, and T. Vekua, Spontaneous increase of magnetic flux and chiral-current reversal in bosonic ladders: swimming against the tide, Phys. Rev. Lett. \textbf{115}, 190402 (2015).

\bibitem{kolley2015strongly}F. Kolley, M. Piraud, I. P. McCulloch, U. Schollw$\ddot{\text{o}}$ck, and F. Heidrich-Meisner, Strongly interacting bosons on a three-leg ladder in the presence of a homogeneous flux, New Journ. of Phys. \textbf{17}, 092001 (2015).

\bibitem{uchino2016analytical}S. Uchino, Analytical approach to a bosonic ladder subject to a magnetic field, Phys. Rev. A \textbf{93}, 053629  (2016).

\bibitem{greschner2016symmetry}S. Greschner, M. Piraud, F. Heidrich-Meisner, I. P. McCulloch, U. Schollw$\ddot{\text{o}}$ck, and T. Vekua, Symmetry-broken states in a system of interacting bosons on a two-leg ladder with a uniform Abelian gauge field, Phys. Rev. A \textbf{94}, 063628 (2016).

\bibitem{natu2015bosons}S. S. Natu, Bosons with long-range interactions on two-leg ladders in artificial magnetic fields, Phys. Rev. A \textbf{92}, 053623 (2015).

\bibitem{zhou2017interaction}X. Zhou, J.-S. Pan, W. Yi, G. Chen, and S. Jia, Interaction-induced exotic vortex states in an optical lattice clock with spin-orbit coupling, Phys. Rev. A \textbf{96}, 023627 (2017).



\bibitem{citro2018quantum} R. Citro, S. De Palo, M. Di Dio, and E. Orignac, Quantum phase transitions of a two-leg bosonic ladder in an artificial gauge field, Phys. Rev. B \textbf{97}, 174523 (2018).


\bibitem{buser2019finite} M. Buser, F. Heidrich-Meisner, and U. Schollw$\ddot{\text{o}}$ck, Finite-temperature properties of interacting bosons on a two-leg flux ladder, Phys. Rev. A \textbf{99}, 053601 (2019).

\bibitem{buser2020interacting} M. Buser, C. Hubig, U. Schollw$\ddot{\text{o}}$ck, L. Tarruell, and F. Heidrich-Meisner, Interacting bosonic flux ladders with a synthetic dimension: ground-state phases and quantum quench dynamics, Phys. Rev. A \textbf{102}, 053314 (2020).

\bibitem{qiao2021quantum} X. Qiao, X.-B. Zhang, Y. Jian, A.-X. Zhang, Z.-F. Yu, and J.-K. Xue, Quantum phases of interacting bosons on biased two-leg ladders with magnetic flux, Phys. Rev. A \textbf{104}, 053323 (2021).

\bibitem{palm2021bosonic} F. A. Palm, M. Buser, J. L\`{e}onard, M. Aidelsburger, U. Schollw$\ddot{\text{o}}$ck, and F. Grusdt, Bosonic Pfaffian state in the Hofstadter-Bose-Hubbard model, Phys. Rev. B \textbf{103}, L161101 (2021).

\bibitem{zheng2023two} Y. Zheng, and S.-J. Yang, Two-body bound and edge bound states in a ladder lattice with synthetic flux, J. Phys. B: At. Mol. Opt. Phys. \textbf{56} 125301 (2023).

\bibitem{huang2024spatial} W. Huang, and Y. Yao, Spatial inversion symmetry breaking of vortex current in a biased-ladder superfluid, Phys. Rev. Research \textbf{6}, 013037 (2024).

\bibitem{giri2023flux} M. K. Giri, B. Paul, and T. Mishra, Flux-induced reentrant dynamics in the quantum walk of interacting bosons, Phys. Rev. A \textbf{108}, 063319 (2023).

\bibitem{giri2024flux} M. K. Giri, B. Paul, and T. Mishra, Flux-enhanced localization and reentrant delocalization in the quench dynamics of two interacting bosons on a Bose-Hubbard ladder, Phys. Rev. A \textbf{109}, 043308 (2024).




\bibitem{maffei2018topological}M. Maffei, A. Dauphin, F. Cardano, M. Lewenstein, and P. Massignan, Topological characterization of chiral models through their long time dynamics, New Journ. of Phys. \textbf{20}  013023 (2018).


\bibitem{initialstate}{These initial states can be achieved by preparing the BEC or two-particle state as the ground state of a free  square flux ladder with a strong local trapping potential for the central two sites, given by $V_{\text{loc}}=-\Gamma (\hat{a}_{0}^\dagger\hat{a}_{0}+\hat{b}_{0}^\dagger\hat{b}_{0})$ with $\Gamma\gg J$. Under these circumstances, the model is simplified into $H_{\text{eff}}\approx-J(\hat{a}^\dagger_{0}\hat{b}_{0}+H.c.)$. The single-particle ground state is just $|\psi_{\text{ini}}\rangle$. Subsequently, the system undergoes quench dynamics by abruptly switching off the local potential and activating the interaction terms.}


\bibitem{SM} See Supplemental Material at [\url{https://journals.aps.org/pra/supplemental/10.1103/w8dc-n153/IICTI_supp_final.pdf}] for more details on the effective model for two-particle bound states, and more discussions on the reversal of chiral transport induced by on-rung interactions, which includes Refs.~\cite{takahashi1997half,qin2014statistics,ke2017multiparticle}.

\bibitem{takahashi1997half} M. Takahashi,  Half-filled Hubbard model at low temperature, Journ. of Phys. C: Sol. Stat. Phys. \textbf{10}, 1289 (1997).

\bibitem{qin2014statistics} X. Qin, Y. Ke, X. Guan, Z. Li, N. Andrei, and C. Lee, Statistics-dependent quantum co-walking of two particles in one-dimensional lattices with nearest-neighbor interactions, Phys. Rev. A \textbf{90}, 062301 (2014).

\bibitem{ke2017multiparticle} Y. Ke, X. Qin, Y. S. Kivshar, and C. Lee, Multiparticle Wannier states and Thouless pumping of interacting bosons, Phys. Rev. A \textbf{95}, 063630 (2017).


\bibitem{vidal2000interaction}J. Vidal, B. Doucot, R. Mosseri, and P. Butaud, Interaction induced delocalization for two particles in a periodic potential, Phys. Rev. Lett.  \textbf{85}, 3906 (2000).
\bibitem{vidal2001disorder}J. Vidal, P. Butaud, B. Doucot, and R. Mosseri, Disorder and interactions in Aharonov-Bohm cages, Phys. Rev. B  \textbf{64}, 155306 (2001).



\end{thebibliography}
\end{document}